\documentclass[useAMS,usenatbib]{mn2e}
\usepackage{graphicx}

\def\lesssim{\mathrel{\hbox{\rlap{\hbox{\lower4pt\hbox{$\sim$}}}\hbox{$<$}}}}
\def\gtrsim{\mathrel{\hbox{\rlap{\hbox{\lower4pt\hbox{$\sim$}}}\hbox{$>$}}}}

\title[Sub-mm detection of a high redshift Type 2 QSO]{Sub-mm detection of a high redshift Type 2 QSO}
\author[V. Mainieri et al.]{V.~Mainieri,$^{1}$\thanks{E-mail: vmainieri@mpe.mpg.de} D.~Rigopoulou,$^{2}$ I.~Lehmann,$^{1}$ S.~Scott,$^{3}$ I.~Matute,$^{1}$ O.~Almaini,$^{4}$ \newauthor P.~Tozzi,$^{5}$ G.~Hasinger,$^{1}$ J.S.~Dunlop$^{3}$\\
$^{1}$Max-Planck-Institut f\"ur extraterrestrische Physik,
Giessenbachstra\ss e, PF 1312, 85741 Garching, Germany\\
$^{2}$Astrophysics, Department of Physics, Keble Road, Oxford OX1
3RH\\ $^{3}$Institut for Astronomy, University of Edinburgh, Royal
Observatory, Blackford Hill, Edinburgh EH9 3HJ\\ $^{4}$School of
Physics \& Astronomy, University of Nottingham, University Park,
Nottingham NG7 2RD\\ $^{5}$INAF, Osservatorio Astronomico di Trieste,
via G.B. Tiepolo 11, I-34131, Trieste, Italy}
\begin{document}

\date{Accepted for publication in MNRAS}
                        
\pagerange{\pageref{firstpage}--\pageref{lastpage}} \pubyear{2004}

\maketitle

\label{firstpage}

\begin{abstract}
We report on the first SCUBA detection of a Type 2 QSO at z=3.660 in
the Chandra Deep Field South. This source is X-ray absorbed, shows
only narrow emission lines in the optical spectrum and is detected in
the sub-mm: it is the ideal candidate in an evolution scheme for AGN
(e.g. Fabian (1999); Page et al. (2004)) of an early phase
corresponding to the main growth of the host galaxy and formation of
the central black hole. The overall photometry (from the radio to the
X-ray energy band) of this source is well reproduced by the spectral
energy distribution (SED) of NGC 6240, while it is incompatible with
the spectrum of a Type 1 QSO (3C273) or a starburst galaxy (Arp
220). Its sub-mm (850 $\mu$m) to X-ray (2 keV) spectral slope
($\alpha_{\rm SX}$) is close to the predicted value for a
Compton-thick AGN in which only $1\%$ of the nuclear emission emerges
through scattering. Using the observed flux at 850 $\mu$m we have
derived a SFR=550--680 M$_{\odot}$/yr and an estimate of the dust
mass, M$_{\rm dust}=4.2 \times 10^8$ M$_{\odot}$.
\end{abstract}


\section{Introduction}

The presence of massive black holes (MBHs) at the centers of most
galaxies appears by now firmly established. A number of correlations
have been observed between the masses of MBHs and the properties of
the galactic bulges hosting them (e.g. \cite{kormendy00},
\cite{merritt01}). Since these correlations extend well beyond the
direct dynamical influence of the MBH, it seems likely that a close
link exists between the growth of the MBH and the formation of their
host galactic bulges.

If these two events are co-eval then we would expect that powerful
black holes are hosted in the centers of the most massive
galaxies. SCUBA galaxies show a strong redshift clustering thus
indicating that they are hosted by very high mass halos
(\cite{blain03}). This is confirmed by CO molecular gas emission line
widths and possible rotation curves (\cite{frayer98}; \cite{genzel03};
\cite{neri03}). The combination of deep X-ray observations and
spectroscopic follow-up campaigns of SCUBA sources
(\cite{alexander04}) led to the discovery that at least $38
\%$ of the SCUBA sources host an AGN, although in the majority of the
cases the contribution of the nucleus is not bolometrically
significant (i.e., $< 20 \%$).\\ A possible scenario would predict
that the black hole grows within a SCUBA galaxy until the obscuring
gas is blown away revealing an AGN (e.g. \cite{fabian99}). A similar
evolution argument has been proposed for ULIRGs (\cite{sanders96}),
which can be considered the closest local analogs to SCUBA
galaxies. We have started to investigate this scenario using a unique
sample of heavily X-ray absorbed (log(N$_{\rm H}) > 22$ cm$^{-2}$) and
very luminous (log(L$_{\rm X}) > 44$ erg s$^{-1}$) AGN, the so-called
Type 2 QSOs, located in the Chandra Deep Field-South (CDF-S) and
selected from
\cite{szokoly04}. In this letter we report on the first sub-mm
detection of a high redshift (z=3.660) Type 2 QSO and discuss the
implications for its nature. We will discuss in a future paper the
properties of the full sample of Type 2 QSOs once our observational
programme is completed. \\ We assume a cosmology with parameters
$\Omega_{\rm M}=0.3$, $\Omega_\Lambda=0.7$ and H$_0=70$ km s$^{-1}$
Mpc$^{-1}$.

\section{Sub-mm observation}

The Submillimetre Common User Bolometer Array (SCUBA,
\cite{holland99}) was employed in photometry mode, with the wide
850:450 filter set, and a standard 60 arcsec chop in azimuth at
7.8Hz. The source was placed in the central bolometer (H7) and the
median of the remaining bolometers was used for additional sky
removal. Flux calibration was performed using Uranus. Telescope
pointing was checked frequently while sky opacity was monitored via
regular skydips at 850 and 450$\mu$m and continuosly via the JCMT
Water Vapour Monitor and the CSO Tau Meter. $\tau _{225}$ ranged from
0.05 to 0.13. The data were reduced using the SURF (SCUBA User
Reduction Facility) software package.  A linear interpolation between
consecutive skydip values was used, and the nearest calibration source
in time applied to account for the gain. After sky subtraction, the
bolometer time-stream was clipped at the 3-$\sigma$ level. Each of the
individual datasets have been tested for consistency with one another
using a Kolmogorov-Smirnov test, rejecting anything below the 5\%
mark.

\begin{figure}
\begin{center}
\includegraphics[width=\columnwidth]{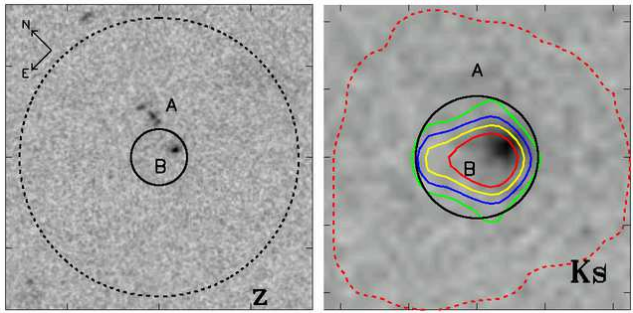}
\end{center}
\caption{{\it Left}: cutout in the z (F850LP) {\it ACS} band of $11\arcsec$
across. The big dashed circle is the 5-arcsec radius from which the
850 $\mu m$ flux can be originated, while the smaller continuum circle
in the center indicates the {\it Chandra} 3 $\sigma$ positional
error. {\it Right}: a zoom of the central part ($5\arcsec$ across) in
the {\it ISAAC}~ K$_{\rm s}$ band, the isointensity X-ray contours at
3, 11, 12, 13, 14 $\sigma$ above the local background are shown. }
\label{figure:fig_acs263}
\end{figure}

We have a firm detection (S/N$\sim 4$) for CDFS-263\footnote{XID from
\cite{giacconi02}} (RA$=3:32:18.86$ DEC$=-27:51:35.5$, z$=3.660$) :
S$_{850}=4.8\pm1.1$ mJy. Since we already know the position of the
X-ray source to sub-arcsec accuracy, the $\sim 14$ arcsec FWHM SCUBA
beam would imply that the sub-mm emission is either coming from this
source or something else within a radius of $\sim 5$ arcsec. Inside
this area we find two optical counterparts in a deep ACS z band image
centered on CDFS-263 (see left panel of
Fig. \ref{figure:fig_acs263}). The X-ray contours clearly point to the
counterpart named 'B' in Fig. \ref{figure:fig_acs263} (the continuum
circle shows the $3\sigma$ {\it Chandra} positional error from
\cite{giacconi02}) . The other possible sub-mm counterpart 'A' has a
complex sub-arcsec structure, where all the different components have
the same optical/near-IR colours therefore likely are at the same
redshift. Since a good spectroscopic redshift is not available for
this object, we have derived a photometric redshift of z$_{\rm
phot}=1.6\pm 0.1$ using the photometry in seven optical/near-IR bands
(Table 3) and the public available code BPZ of \cite{benitez00} (see
\cite{mainieri04} for details on the procedure).  Assuming this
value, source 'A' has to have L$_{\rm FIR}
\approx 10^{11.5}$ to shine in the sub-mm, and the probability to have
by chance such a powerful source in the SCUBA error circle is low. The
same area of the sky imaged in the K$_{\rm S}$ band with ISAAC/VLT
shows only the 'B' counterpart (right panel of Fig.
\ref{figure:fig_acs263}) reinforcing its chances to be the real sub-mm
source.

\begin{figure}
\begin{center}
\includegraphics[angle=-90,width=\columnwidth]{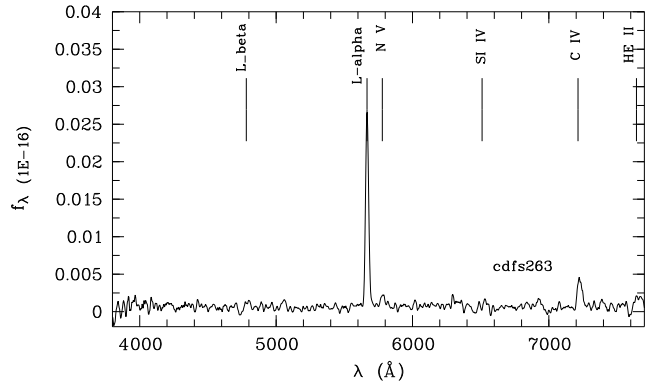}
\end{center}
\caption{VLT FORS spectrum of CDFS-263 (z$=3.660$) from Szokoly et al. (2004).  }
\label{figure:fig_opt_spec}
\end{figure}

\begin{table}
\begin{center}
\caption{Optical emission line properties of CDFS-263. FWHM and EW 
are rest frame}
\begin{tabular}{lccc}
\noalign{\smallskip \hrule \smallskip}
line & flux & FWHM & EW\\ & $[10^{-18}$ erg cm$^{-2}$ s$^{-1}]$ &
$[$km s$^{-1}]$ & $[$\AA$]$ \\
\noalign{\smallskip \hrule \smallskip}
Ly$\alpha$ $\lambda1216$   & 2.96$\pm$0.04  & 1000$\pm$10   & 29.6$\pm$3.0 \\
N V $\lambda1240$         & 0.17$\pm$0.03  & 1580$\pm$140   & 2.4$\pm$0.6  \\
%SI IV  $\lambda1397$      &  -             &   -         &             \\
C IV $\lambda1548$        & 0.40$\pm$0.03  & 1400$\pm$50   & 5.7$\pm$0.9 \\
\noalign{\smallskip \hrule \smallskip}
\end{tabular}
\end{center}
\label{table:lines}
\end{table}

\section{Optical and X-ray spectroscopy}

To help assign a reliable classification of CDFS-263 as a Type 2 quasar
we have determined its optical emission line and X-ray spectral
properties. 

VLT FORS1 spectra of CDFS-263 were taken in 2000 during the
spectroscopic identification programme of the CDF-S field
(\cite{szokoly04}).  The object was observed with a multi-slit mask
using 1.2 arcsec wide single slits and grism 150I with a spectral
dispersion of $\sim5.5$ \AA/pixel, which corresponds to a spectral
resolution of $\sim 11$ \AA. The seeing was about 0.5--0.6 arcsec. The
total integrated exposure time of the spectrum (see
Fig. \ref{figure:fig_opt_spec}) is about 3.3 hours. The measured
redshift is $z=3.660 \pm 0.005$.
 
The optical spectrum of CDFS-263 shows a strong
Ly$\alpha$$\lambda1216$ emission line and the two fainter lines of N V
$\lambda1240$ and C IV $\lambda1548$, whereas no further lines as
e.g. Ly$\beta$$\lambda1026$ or SI IV $\lambda1397$ are visible. Each
emission line was fitted with a single Gaussian profile applying the
Levenberg-Marquardt algorithm (\cite{press92}). The four adjustable
parameters are the total line flux, the central wavelength, the FWHM
in \AA ~and the flux of a local linear continuum. From this set of
parameters we have calculated the FWHM in km s$^{-1}$ (corrected for
instrumental resolution) and the rest frame EW in \AA. The line flux,
the rest frame FWHM and the rest frame EW of the visible lines are
given in Table 1.

\begin{table}
\begin{center}
\caption{Best fit spectral model parameters for CDFS 263 (9.3596E+05 sec)
         Galactic N$_{H}=9.13\times 10^{19}$ cm$^{-2}$. If we include
         the galactic N$_{H}$ there is only a marginal change of the
         intrinsic N$_{H}$ value.}
\begin{tabular}{lcc}
\noalign{\smallskip \hrule \smallskip}
parameter & \multicolumn{2}{c}{PL $+$ intrinsic absorption }\\
          & model1 &  model 2\\
\noalign{\smallskip \hrule \smallskip}
N$_{H}$ (10$^{22}$ cm$^{-2}$)  &  0$^{+140}_{-0}$      & 88$^{+95}_{-75}$\\
z                              & \multicolumn{2}{c}{3.660 (froozen)}\\
$\Gamma$                       &  0.34$^{+2.59}_{-0.59}$ &1.90 (frozen)      \\
$F_{\rm p15}$                  &  1.04$^{+0.66}_{-0.52}$ & 1.71$^{+1.37}_{-3.42}
$    \\
$\chi^{2}$/dof                 &  1.19/4     &  3.37/5       \\
f$_{0.5-10.0}$ (absorbed)      &  1.93E-15   &  1.04E-15      \\
f$_{0.5-10.0}$ (unabsorbed)     &  -          &  2.00E-15      \\
L$_{0.5-10.0}$ (absorbed)       &  0.19E-44   &  0.37E-44      \\
L$_{0.5-10.0}$ (unabsorbed)     &  -          &  7.64E-44      \\
\noalign{\smallskip \hrule \smallskip}
\end{tabular}
\end{center}
\label{table:tab_xray}
Fit parameters are shown with 90\% errors ($\Delta \chi^2=2.7$).  The
model and parameter definitions are: Model Components-- {\bf PL}:
Power-law with a photon index of $\Gamma$ and a 0.5-7 keV flux of
$F_{\rm p15}$ in units of 10$^{-15}$ erg cm$^{-2}$ s$^{-1}$ using the
XSPEC model {\it pegpwrlw}.  {\bf Intrinsic absorption:}
Photo-electric absorption using Wisconsin cross-sections {\it zwabs}
(Morrison \& McCammon 1983), where N$_{H}$ is the equivalent Hydrogen
column in units of 10$^{22}$ atoms cm$^{-2}$ and z is the redshift.\\
The luminosities reported in the table are in the rest frame [0.5-10]
keV energy band.
\end{table}

N V $\lambda1240$ is observed at the 4$\sigma$ level. The line fluxes
are a factor of $\sim20--100$ lower compared to those of the prototype
Type 2 QSO CDFS-202 (\cite{norman02}). The emission line widths are
narrow with a FWHM of $\lesssim 1500$ km s$^{-1}$, whereas the mean line
width of the permitted lines for X-ray selected Seyfert 1 and QSO1
ranges from $\sim3000$ to $\sim5000$ km s$^{-1}$ (see
\cite{lehmann01}). The narrow line widths and the observed 0.5-10 keV
luminosity of $\sim1.6 \times 10^{44}$ erg s$^{-1}$ indicate a Type 2
QSO (\cite{szokoly04}).

The analysis of the X-ray spectrum of CDFS-263 is essential to
determine its amount of intrinsic X-ray absorption and its unabsorbed
X-ray luminosity.  We have used the public {\it Chandra} ACIS-I data
of the 1 Msec CDF-S survey (\cite{giacconi02}). The total integrated
exposure time is about 936 ksec for CDFS-263.  The ACIS-I spectrum of
the source was extracted from a circular region of $\sim 5$ arcsec in
radius and the background spectrum was obtained from a source free
region. The source spectrum contains 120 counts. The spectra were
grouped to have at least 10 photons per bin.

\begin{figure}
\begin{center}
\includegraphics[angle=-90,width=\columnwidth]{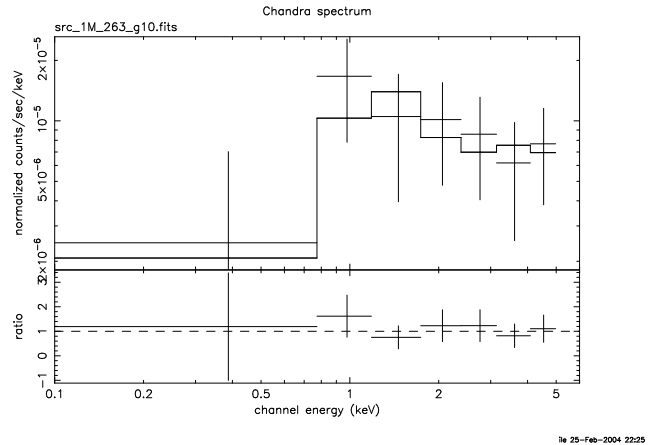}
\end{center}
\caption{The observed Chandra ACIS-I spectrum of CDFS 263 and the best-fit 
absorbed powerlaw model (2). The spectrum was grupped to have at least
10 photons per bin. The bottom panel shows the ratios between the
observed data points (crosses) and the best-fit model data points
(continuum line).  }
\label{figure:fig_xray_spec}
\end{figure}

Due to the small number of source photons we can only get an estimate
of the X-ray spectral properties. The spectrum was analysed using
XSPEC 11.3.  A simple absorbed powerlaw model (1 in Table 2) gives a
reasonable fit. The photon index of $\Gamma=0.34$ is low compared to
that of $\Gamma\sim1.9$ found for most Seyfert galaxies and
quasars. This points to a large intrinsic absorption of
CDFS-263. Therefore we have frozen the photon index to 1.9 (model
2). The results of both models are shown in Table 2. All the errors
quoted we calculated at 90\% confidence level for two interesting
parameters. As expected, the best fit model 2 shows a large N$_{H}$ of
$\sim10^{24}$ cm$^{-2}$, which implies nearly Compton-thick
absorption. If we correct for intrinsic absorption the 0.5-10 keV rest
frame luminosity of CDFS-263 increases by a factor of $\sim 20$ (see
Table 2). The best fit model (2) and the X-ray data points are shown
in Fig. \ref{figure:fig_xray_spec}. Whereas the quality of the ACIS-I
spectrum is low, the strong intrinsic absorption of CDFS-263 (at least
N$_{H}>1 \times 10^{23}$ cm$^{-2}$) is fairly well constrained.

The narrow optical emission lines (FWHM$<1500$ km s$^{-1}$), the large
intrinsic absoprtion (log N$_{H}> 22$) and the large absorption
corrected X-ray luminosity (log L$_{X} > 44$) in the 0.5-10 keV energy
band favour the Type 2 quasar nature of CDFS-263.

\section{Discussion}

\subsection{Spectral energy distribution}

\begin{figure}
\begin{center}
\includegraphics[width=\columnwidth]{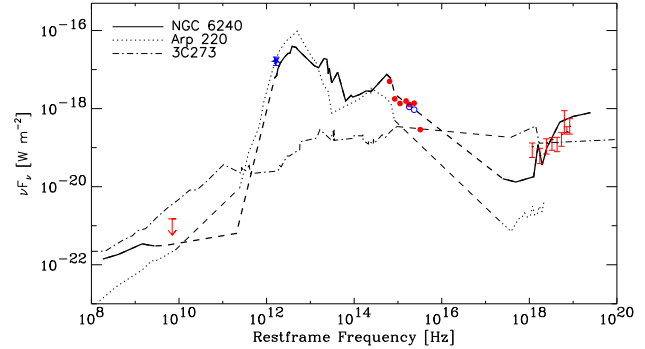} \\
\end{center}
\caption{Detections and upper limits for CDFS-263 plotted over the SED 
of NGC 6240 ({\it solid}), Arp 220 ({\it dotted}) and 3C273 ({\it
dash-dotted}). The dashed line shows where the photometry is not
available and template is extrapolated. The CDFS-263 data have been
shifted to the rest frame and the empty circles show the line
subtracted magnitudes (see text).}
\label{figure:fig_sed}
\end{figure}

CDFS-263 has been observed in a wide range of energy bands. Photometry
of extraordinary quality in the BVIz filters has been obtained with
the ACS camera on board of HST as part of the GOODS survey
(\cite{giavalisco04})\footnote{These data are publicly available at:
http://www.stsci.edu/science/goods/}. As part of the same survey, ESO
carried out deep near-IR imaging with ISAAC/VLT in J, H and K$_{\rm
s}$ bands\footnote{These data are publicly available at:
http://www.eso.org/science/goods/}. Finally, we have a radio upper
limit from VLA observations at 20 cm of F$_\nu < 100$ $\mu$Jy
(K. Kellerman, private comunication). The available photometry for
CDFS-263 in summarized in Table 3.\\ According to the radio-power
criterion given by
\cite{stocke92}, we can classify this AGN as radio-quiet. Hence, 
a non-thermal (synchrotron) contribution to the sub-mm flux is
expected to be small (\cite{carleton87}; \cite{hughes93}).\\We compare
the photometry of CDFS-263 with the SED of several well-studied local
sources. We choose 3C273 as a template of an unabsorbed quasar, NGC
6240 a well-studied (U)LIRG which hosts a Compton-thick AGN, and Arp
220 as a starburst galaxy where X-ray binaries are the major source of
its X-ray emission (\cite{iwasawa01}). Due to the difference in
redshift between our source and these templates, we have photometry
for CDFS-263 in wavelength regions not covered for the local
sources. In particular for CDFS-263, the {\it V} filter contains
Ly$\alpha$ while the {\it i} band contains the CIV line. We have
subtracted the contribution of these lines in the observed fluxes and
compared them with the templates SED in Fig. \ref{figure:fig_sed}. The
wide-band energy distribution of CDFS-263 is inconsistent with the
spectrum of an unabsorbed QSO (dash-dotted line in
Fig. \ref{figure:fig_sed}): if we anchor 3C273 to the X-ray
observations, we underpredict the fluxes at the other
wavelengths. Using the SED of NGC 6240, we can reproduce in a
consistent way the observed sub-mm, optical, near-IR and X-ray fluxes
of our source (solid thick line in Fig. \ref{figure:fig_sed}). The
observed flux in the B band is low compared with the expected one,
this can be ascribed to absorption due to the Ly$\alpha$ forest for a
source at $z=3.660$ (see also \cite{stern02}).  Finally, the energy
distribution of CDFS-263 does not resemble the SED of Arp 220 (dotted
line in Fig. \ref{figure:fig_sed}): if we fix the template at the
observed sub-mm flux we underestimate the X-ray flux by two orders of
magnitudes. We note that Arp 220 appears to be an order of magnitude
underluminous in soft X-rays, given its FIR luminosity, if compared
with other starbursts galaxies (see Fig. 3 of
\cite{iwasawa01}), but this is not enough to reconcile the
photometry of CDFS-263 with a starburst template.\\

\begin{table}
\begin{center}
\caption{Photometry of CDFS-263.}
\begin{tabular}{cccc}
\noalign{\smallskip \hrule \smallskip}
Observed & Rest-frame & Observed & $\nu$F$_{\nu}$ \\ Band & Band & AB
Magnitude & [W/m$^2$] \\
\noalign{\smallskip \hrule \smallskip}
 2-10 keV $^a$ & 27.9 keV & ... &  2.6E-19 \\
\smallskip
 0.5-2 keV $^a$ & 5.8 keV & ... &  8.1E-20 \\
\smallskip
 F435W $^b$ &  927\AA & 27.7$\pm$0.4 &  2.9E-19 \\
\smallskip
 F606W  $^b$ &  1277\AA & 25.67$\pm$0.05 &  1.4E-18  \\
\smallskip
 F775W $^b$ &  1654\AA & 25.38$\pm$0.08 &  1.3E-18 \\
\smallskip
 F850LP $^b$ &  1957\AA & 25.08$\pm$0.08 &  1.6E-18  \\
\smallskip
 J $^c$ &  2690\AA & 24.5$\pm$0.1 &  1.4E-18 \\
\smallskip
 H $^d$ &  3550\AA & 23.9$\pm$0.4 &  1.8E-18 \\
\smallskip
 K$_{\rm s}$ $^c$ &  4660\AA & 22.46$\pm$0.04 &  5.0E-18  \\
\smallskip
 850$\mu$m $^e$ &  1.6E3 GHz & ... &  1.6E-17 \\
\smallskip
 20 cm $^f$ &  6.9 GHz & ... &  $<$1.5E-21 \\
 
\noalign{\smallskip \hrule \smallskip}
\end{tabular}
\end{center}
\label{table:table_CDFS263}
$^a${\it Chandra}/ACIS; $^b${\it HST}/ACS; $^c${\it VLT}/ISAAC; $^d${\it NTT}/SOFI; $^e${\it JCMT}/SCUBA; $^f${\it VLA}
\end{table}

\subsection{Sub-mm to X-ray spectral index}

\begin{figure}
\begin{center}
\includegraphics[angle=0,width=\columnwidth]{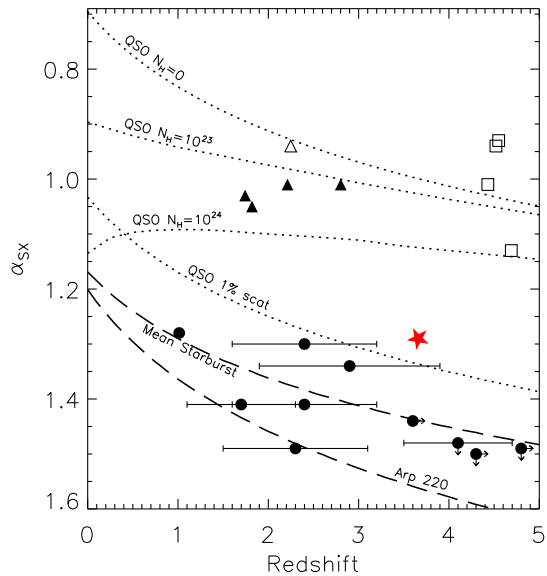}
\end{center}
\caption{Sub-mm to X-ray spectral index ($\alpha_{\rm SX}$) as a function 
of redshift. The dotted and dashed lines are the expected values of
$\alpha_{\rm SX}$ for a set of SEDs (see text). The values for
CDFS-263 (star), the sample from Alexander et al. (2003) (circles),
Page et al. (2001) (filled triangles), Page et al. (2004) (empty
triangle) and a sample of Type 1 QSOs at $z>4$ (squares) are shown.}
\label{figure:fig_alphaSX}
\end{figure}

To further compare the properties of our source with local templates
and other SCUBA observations of X-ray sources, we derive the sub-mm
(850 $\mu$m) to X-ray (2 keV) spectral slope ($\alpha_{\rm SX}$) of
CDFS-263. Fig. \ref{figure:fig_alphaSX} shows the expected values of
$\alpha_{\rm SX}$, as a function of redshift, for a number of template
SEDs. Namely, a quasar with no absorption in the X-ray or with
different amounts of absorption (N$_{\rm H}=10^{23},10^{24}$ and
Compton-thick with 1$\%$ of the nuclear emission scattered); a mean
starburst template and Arp 220. These templates are described in
detail in \cite{almaini03}.\\ CDFS-263 yields a value of $\alpha_{\rm
SX}=1.29$ (star symbol in Fig. \ref{figure:fig_alphaSX}). This value
is clearly incompatible with an unabsorbed AGN, as we already expected
from the optical and X-ray spectral properties of the source. On the
other hand, it is close to the spectral index of a Compton-thick AGN
in which only $1\%$ of the nuclear emission emerges through
scattering. The X-ray spectral fit yields a column density of $\sim
10^{24}$ cm$^{-2}$ (see Table 2). One way to obtain this value of
$\alpha_{\rm SX}$ is a sub-mm flux due to starburst activity plus an
absorbed AGN in the center of the host galaxy that accounts for the
X-ray emission, which is higher than that expected for a starburst
galaxy alone (dashed line in Fig. \ref{figure:fig_alphaSX}) (see also
\cite{fabian00}; \cite{alexander03}; \cite{almaini03}). For 
comparison we report in Fig. \ref{figure:fig_alphaSX} other X-ray
selected sources with sub-mm data measurements from the
literature. Namely, the CDF-N sources from \cite{alexander03}
(circles) the majority of which show $\alpha_{\rm SX}$ values higher
than CDFS-263, and compatible with the expected indices for a
starburst galaxy, although all except one of them lack a precise
redshift measurement. The X-ray absorbed (filled triangles) and X-ray
unabsorbed (empty triangle) ROSAT BLAGN from
\cite{page01} and \cite{page04} are sampling  a different region in 
Fig. \ref{figure:fig_alphaSX}, with their $\alpha_{\rm SX}$ values
being closer to pure QSO tracks in agreement with their nature as
unabsorbed AGN in the optical band. Finally, we show also the
$\alpha_{\rm SX}$ values of some $z>4$ Type 1 QSOs (square) with X-ray
(\cite{kaspi00}) and sub-mm (\cite{mcmahon99}) observations. The
location of our source in the $\alpha_{\rm SX}$ vs $z$ plot is
peculiar: it has a lower $\alpha_{\rm SX}$ than any source from
\cite{alexander03} at high redshift and a higher value of the BLAGN
from \cite{page01} \&
\cite{page04} and the Type 1 QSOs at $z>4$.

\subsection{Star Formation Rate, dust mass and black hole mass}

Under the hypothesis that the sub-mm flux is due to star formation
activity, we estimate, using the measured flux at 850 $\mu$m, the
dust-enshrouded Star Formation Rate (SFR) and the dust mass. We note
that the temperature of the dust heated by the AGN would be too high
($\gg 100$ K) to emit in the sub-mm. The far-infrared luminosity
(L$_{FIR}$) provides a reasonable measure of the active formation of
massive stars.  Assuming a dust temperature of 40K and an emissivity
index $\beta$ = 1.2 (e.g. \cite{dunne00}) we interpolate from the 850
$\mu$m luminosity and get an estimate of the 60 $\mu$m luminosity
assuming an optically-thin greybody. The SFR is given by:

$SFR = \epsilon 10 ^{-10} L_{60}/L_{sun} (M_{\odot}/yr)$

\noindent where $\epsilon$ is the fraction of optical$/$UV 
light being radiated in the far-infrared and depends on the IMF and
the fraction of the optical$/$UV light that is absorbed and
re-radiated in the FIR. For $\epsilon = 0.6--2$ we get $SFR = 550 --
680 M_{\odot}/yr$ (assuming $\Omega_{\lambda} = 0.7$ and H$_0 = 70$).

\noindent The dust mass is given by:

M$_{dust} = 1/ (1+z) S_{850} D_{L} ^2 / k_{d(rest)} B(v_{rest}, T_d)$

\noindent where z is the redshift, S$_{850}$ is the observed 
flux density, D$_L$ is the luminosity distance, k$_{rest}$ is the rest
frequency absorption coefficient and B is the Planck function. Using
K$_{rest} = 0.15$ (see \cite{scott02}) we derive an M$_{dust} =
4.2\times 10^8$ M$_{\odot}$.  The dust mass estimate assumes optically
thin thermal emission with no contribution from bremsstrahlung or
synchrotron radiation. 

Finally, we compute the rest frame absorption corrected L$_{\rm
X}[2-10 $ keV$]=4.5 \times 10^{44}$ erg s$^{-1}$ and using the
relation given by \cite{marconi04} estimate the bolometric luminosity,
L$_{\rm bol}=2.4 \times 10^{46}$ erg s$^{-1}$. Assuming that L$_{\rm
bol}=$L$_{\rm Edd}$, where L$_{\rm Edd}$ is the Eddington luminosity,
we derive a lower limit on the black hole mass: M$_{\rm BH}=1.9 \times
10^8$ M$_{\odot}$.

\section{Conclusions}

We have a firm ($4 \sigma$) SCUBA detection of a Type 2 QSO at
redshift $z=3.660$ (CDFS-263). Its SED is consistent, from radio to
X-ray wavelengths, with the shape of NGC 6240 while it can not be
reproduced with the SED of 3C273 or Arp 220. Futhermore, we derive an
$\alpha_{\rm SX}=1.29$. We conclude that CDFS-263 can be powered by a
Compton thick AGN (we found N$_{\rm H}\sim 1-20 \times 10^{23}$
cm$^{-2}$). 

CDFS-263 is a strong candidate for an AGN in the initial phase (before
the ``X-ray absorbed phase'') described by \cite{page04} that
corresponds to the main growth period of the host galaxy spheroid. It
is a luminous X-ray source (L$_{\rm X}>10^{44}$ erg s$^{-1}$) with
high X-ray absorption (N$_{\rm H} > 10^{23}$ cm$^{-2}$) and detected
in the sub-mm, S$_{850}=4.8 \pm 1.1$ mJy.

\section*{Acknowledgments}

We thank Kazushi Iwasawa for providing us the Beppo-SAX data points
for Arp 220 and Andrea Merloni for useful discussions. We thank the
referee for helpful comments that improved the manuscript.\\ We
acknowledge support from the European Community RTN Network ``POE''
(under contract HPRN-CT-2000-00138).

\label{lastpage} 

\begin{thebibliography}{99}

\bibitem[\protect\citeauthoryear{Alexander et al.}{2003}]{alexander03}
Alexander D.M. et al., 2003, ApJ, 125, 383
\bibitem[\protect\citeauthoryear{Alexander et al.}{2004}]{alexander04}
Alexander D.M. et al., 2004, Proceedings of the ESO/USM/MPE Workshop
on ``Multiwavelength Mapping of Galaxy Formation and Evolution'',
eds. R.Bender and A.Renzini [astro-ph/0401129]
\bibitem[\protect\citeauthoryear{Almaini et al.}{2003}]{almaini03}
Almaini D.M et al., 2003, MNRAS, 338, 303
\bibitem[\protect\citeauthoryear{Archibald et al.}{2001}]{archibald01}
Archibald E.N. et al., 2001, MNRAS, 323, 417
\bibitem[\protect\citeauthoryear{Benitez}{2000}]{benitez00}
Benitez N. 2000, ApJ, 536, 2000
\bibitem[\protect\citeauthoryear{Blain et al.}{2003}]{blain03}
Blain A.W. et al., 2003, ApJ, 611, 725
\bibitem[\protect\citeauthoryear{Borys et al.}{2003}]{borys03} Borys
C. et al., 2003, MNRAS, 344, 385
\bibitem[\protect\citeauthoryear{Carleton et al.}{1987}]{carleton87}
Carleton N.P. et al., 1987, ApJ,318, 595
\bibitem[\protect\citeauthoryear{Chapman et al.}{2003}]{chapman03}
Chapman S.C., Blain A.W., Ivison R.J., Smail I.R., 2003, Nat, 422, 695
\bibitem[\protect\citeauthoryear{Dunne et al.}{2000}]{dunne00}
Dunne L., Eales S., Edmunds M. et al., 2000, MNRAS, 315, 115
\bibitem[\protect\citeauthoryear{Fabian }{1999}]{fabian99}
Fabian A.C., 1999, MNRAS, 308, L39
\bibitem[\protect\citeauthoryear{Fabian et al.}{2000}]{fabian00}
Fabian A.C. et al., 2000, MNRAS, 315, L8
\bibitem[\protect\citeauthoryear{Frayer et al.}{1998}]{frayer98}
Frayer D. et al., 1998, ApJ, 506, L7
\bibitem[\protect\citeauthoryear{Genzel et al.}{2003}]{genzel03}
Genzel R. et al., 2003, ApJ, 584, 633
\bibitem[\protect\citeauthoryear{Giacconi et al.}{2002}]{giacconi02}
Giacconi R. et al. 2002, ApJS, 139, 369
\bibitem[\protect\citeauthoryear{Giavalisco et al.}{2004}]{giavalisco04}
Giavalisco M. et al., 2004, ApJ, 600, 93
\bibitem[\protect\citeauthoryear{Holland et al.}{1999}]{holland99}
Holland W.S. et al., 1999, MNRAS, 303, 659
\bibitem[\protect\citeauthoryear{Hughes et al.}{1993}]{hughes93}
Hughes D.H. et al., 1993, MNRAS, 263, 607
\bibitem[\protect\citeauthoryear{Iwasawa et al.}{2001}]{iwasawa01}
Iwasawa K. et al., 2001, MNRAS, 326, 894
\bibitem[\protect\citeauthoryear{Kaspi et al.}{2000}]{kaspi00}
Kaspi S. et al., 2000, AJ, 119, 2031
\bibitem[\protect\citeauthoryear{Kormendy \&
Gebhardt}{2000}]{kormendy00} Kormendy J. \& Gebhardt K., 2000,
Am. Inst. Phys., p.63 [astro-ph/0105230]
\bibitem[\protect\citeauthoryear{Lehmann et al.}{2001}]{lehmann01}
Lehmann I. et al., 2001, A\&A, 371, 833
\bibitem[\protect\citeauthoryear{Marconi et al.}{2004}]{marconi04}
Marconi A. et al., 2004, MNRAS, 351, 169
\bibitem[\protect\citeauthoryear{Mainieri et al.}{2004}]{mainieri04}
Mainieri V. et al., 2004, submitted to A\&A
\bibitem[\protect\citeauthoryear{McMahon et al.}{1999}]{mcmahon99}
McMahon R.G. et al., 1999, MNRAS, 309, L1
\bibitem[\protect\citeauthoryear{Merritt \&
Ferrarese}{2001}]{merritt01} Merritt D. \& Ferrarese L., 2001, MNRAS,
320, L30
\bibitem[\protect\citeauthoryear{Morrison et al.}{1983}]{morrison83}
Morrison, R., McCammon, D., 1983, ApJ, 270, 199
\bibitem[\protect\citeauthoryear{Neri et al.}{2003}]{neri03}
Neri R. et al., 2003, ApJ, 597, 113
\bibitem[\protect\citeauthoryear{Norman et al.}{2002}]{norman02}
Norman C. et al., 2002, ApJ, 571, 218
\bibitem[\protect\citeauthoryear{Page et al.}{2001}]{page01}
Page M.J. et al., 2001, Science, 294, 2516
\bibitem[\protect\citeauthoryear{Page et al.}{2004}]{page04}
Page M.J. et al., 2004, accepted by ApJL, [astro-ph/0407171]
\bibitem[\protect\citeauthoryear{Press et al.}{1992}]{press92}
Press et al., 1992, Numerical recipes in FORTRAN, Example book, 2nd
edition (Cambridge Press)
\bibitem[\protect\citeauthoryear{Sanders \& Mirabel}{1996}]{sanders96}
Sanders D.B. \& Mirabel I.F., 1996, ARA\&A, 34, 749
\bibitem[\protect\citeauthoryear{Scott et al.}{2002}]{scott02}
Scott S.E., Fox M.J., Dunlop J.S. et al., 2002, MNRAS, 331, 817
\bibitem[\protect\citeauthoryear{Stern et al.}{2002}]{stern02}
Stern D.. et al., 2002, ApJ, 568, 71
\bibitem[\protect\citeauthoryear{Stevens et al.}{2004}]{stevens04}
Stevens J.A., Page M.J., Ivison R.J., et al., 2004, ApJ, 604, 17
\bibitem[\protect\citeauthoryear{Stocke et al.}{1992}]{stocke92}
Stocke J.T. et al., 1992, ApJ, 396, 487
\bibitem[\protect\citeauthoryear{Szokoly et al.}{2004}]{szokoly04}
Szokoly G.P. et al., 2004, ApJS in press, [astro-ph/0312324]

\end{thebibliography}
\end{document}